# Characterizing QoS Parameters and Application of Soft-QoS Scheme for 3G Wireless Networks

Mostafa Zaman Chowdhury, Mohd. Noor Islam, Young Min Seo, Young Ki Lee,
Sang Bum Kang, Sun Woong Choi, and Yeong Min Jang[*].
Kookmin University, Korea.
[*]yjang@kookmin.ac.kr

*Abstract* — In wireless communication systems, Quality of Service (QoS) is one of the most important issues from both the users and operators point of view. All the parameters related to QoS are not same important for all users and applications. The satisfaction level of different users also does not depend on same QoS parameters. In this paper, we discuss the QoS parameters and then propose a priority order of QoS parameters based on protocol layers and service applications. We present the relation among the QoS parameters those influence the performance of other QoS parameters and, finally, we demonstrate the numerical analysis results for our proposed soft-QoS scheme to reduce the dropped call rate which is the most important QoS parameter for all types of services.

## 1. Introduction

Future wireless networks will be heterogeneous. The heterogeneous wireless networks integrate different access networks, such as IEEE 802.15 WPAN, IEEE 802.11 WLAN, IEEE 802.16 WMAN, GPRS/EDGE, cdma2000, WCDMA and satellite network, etc. [1]. QoS provisioning and management is one of the most important issues to provide various services through heterogeneous networks efficiently and economically. As different networks can not provide same QoS level and different services need different QoS levels, provisioning of QoS in an efficient manner is a complex one, involving multiple inter-related aspects, including resource provisioning, call admission control, traffic policing, routing, and pricing. To meet QoS requirements accurately, all the parameters of a service should work properly and network resources must be used more efficiently.

Different solutions already have been proposed to provide QoS throughout the seven layers of OSI reference model. It is believed that a best Quality of Experience (QoE) level can be achieved when all the QoS parameters of all layers can be considered as a whole instead of a single entity. QoS management and policy help to set and evaluate QoS policies and goals.

There are many QoS parameters defined by different standardization bodies like 3GPP, 3GPP2, GSMA, ITU-T, Mobile WiMAX Forum etc. [2~7] to meet user's QoS requirement. They also defined the QoS requirement for different services. Among these QoS parameters, most of them are related to application layer, network layer or physical layer. As the QoS parameters for all the layers are not same, the QoS provisioning and management mechanism for different layers are also different. All these QoS parameters are not equally important from users or operators point of view. Also QoS requirement for different services are different. A QoS parameter priority order is needed to make the QoS management and policy with more easily for better user's satisfaction level within limited resources.

Some QoS parameters performances influence other QoS parameters performances. So performance degradation of one parameter can degrade the performance of some other parameters. Hence knowledge about these QoS parameters relationship is also needed to operate a network with better QoS management system.

Whenever a session starts a call for any types of services, the users always want to complete the session without any interruption. Hence after starting a session, dropping a call for any reason is the most undesirable situation from user point of view. A soft-QoS scheme based on critical bandwidth ratio [8] can be used to reduce the dropped call rate incase of limited bandwidth. Thus can accept more handoff calls when a user move from one cell to other cell and the bandwidth of the target cell is limited at that time. This soft-QoS scheme can also be used to accommodate new calls for medium traffic load condition.

This paper is organized as follows. Section 2 provides a list for the QoS parameters according to priority or importance for different type of services. Three QoS relationship models for voice, video and data services are presented in section 3. These models summarize the relationship among QoS parameters those influence the performances of other parameters. In Section 4, a soft-QoS scheme has been proposed to reduce dropped call rate. The numerical results for the proposed soft-QoS scheme are provided in section 5. Finally, we concluded in section 6.

## 2. QoS Parameters with Priority Order

The QoS requirement and importance vary according to service type, price and user's requirement. Also the QoS provisioning mechanism of a network depends on user's requirement, availability of resources, price, service types etc. Table 1 provides a list that includes most of the QoS parameters which are related to performance measuring parameters of heterogeneous 3G networks. This list divides the QoS parameters according to protocol layers (application, network and physical) and type of services (voice, video and data). The QoS parameters enlisted in the Table 1 are arranged (top to bottom) according to the priority order of the parameters.

This list can be helpful to know about the QoS parameters at a glance. As the demand and user satisfaction level are related to price and also sometime resources are limited, this priority order can be helpful for the optimization of QoS parameters. This priority order is also useful for the operators to monitor the QoS parameters and to manage the networks resources in an efficient way so that they can provide maximum QoE level for the users. Normally the users can realize the effect of QoS parameters those are related to application layer. But the performances of the QoS parameters of the other layers influence the QoS parameters of application layer.

**Table 1. QoS Parameters List with Priority Order**

| Layer / Service | Application Layer | Network Layer | Physical Layer |
|---|---|---|---|
| Voice | <ul><li>Dropped Call Rate</li><li>End-to-End One Way Delay</li><li>Delay Variation, Jitter</li><li>Call Completion Rate</li><li>Speech Quality (MOS, R value)</li><li>Service Accessibility</li><li>Poor Quality Rate within Call Duration Time</li><li>Post Dialing Delay (PDD)</li><li>Call Setup Success Ratio</li><li>Start-up Delay</li><li>Application Response Time</li><li>Frame Erasure Rate</li><li>Codec Delay</li></ul> | <ul><li>Handover Success Ratio</li><li>Delay Variation, Jitter</li><li>Handover Delay</li><li>Max Transfer Delay</li><li>Guaranteed Bit Rate</li><li>Network Accessibility</li><li>Maximum Bit Rate</li><li>Frame Error Rate</li><li>Packet Error Rate</li><li>Packet Loss Rate</li><li>Traffic Handling Priority</li><li>Allocation/Retention Priority</li><li>One Way Radio Access Network Transfer Delay</li></ul> | |
| Video | <ul><li>Dropped Call Rate</li><li>End-to-End One Way Delay</li><li>Lip-synch Delay</li><li>Resolution</li><li>Delay Variation, Jitter</li><li>Application Response Time</li><li>Call Completion Rate</li><li>Service Accessibility</li><li>Packet Error Rate</li><li>Frame Erasure Rate</li><li>Packet Loss Rate</li><li>Frame Per Second (FPS)</li><li>Blockiness</li><li>Blurring</li><li>Jerkiness</li><li>Poor Quality Rate within Call Duration Time</li><li>Post Dialing Delay (PDD)</li><li>Call Setup Success Ratio</li><li>Start-up Delay</li><li>Codec Delay</li><li>Codec Type</li></ul> | <ul><li>Handover Success Ratio</li><li>Max Transfer Delay</li><li>Handover Delay</li><li>Delay Variation, Jitter</li><li>Guaranteed Bit Rate</li><li>Network Accessibility</li><li>Frame Error Rate</li><li>Buffering Capacity and Delay</li><li>Maximum Bit Rate</li><li>Traffic Handling Priority</li><li>Allocation/Retention Priority</li><li>One Way Radio Access Network Transfer Delay</li></ul> | <ul><li>BER</li><li>CINR</li><li>$E_c/I_0$ ($E_b/I_0$)</li><li>SNR</li><li>Minimum Path Loss</li><li>Noise Figure and Noise Power</li><li>Received Signal Strength Indicator (RSSI)</li><li>UE TX Power</li><li>Codeword Decoding Error Probability</li><li>Maximum BTS Power</li><li>Target $E_b/N_0$</li><li>Channel Capacity</li></ul> |
| Data | <ul><li>Dropped Call Rate</li><li>Packet Loss Rate</li><li>Packet Error Rate</li><li>Data Throughput</li><li>Completed Session Ratio Packet Switched Data</li><li>Service Accessibility Rate Packet Switched Data</li><li>Web Browsing One Way Delay</li><li>End-to-End One Way Delay</li><li>Delay Variation, Jitter</li></ul> | <ul><li>Handover Success Ratio</li><li>Frame Error Rate</li><li>Data Throughput</li><li>Network Accessibility</li><li>Maximum Bit Rate</li><li>Handover Delay</li><li>One Way Radio Access Network Transfer Delay</li></ul> | |

## 3. QoS Parameters Relationship

There are some QoS parameters which influence the performance of other QoS parameters. This section proposes three models those summarize these QoS relationships. Three models based on voice, video and data services are shown here. Each of these three models has three parts. They are application layer, physical layer and network layer. The characteristics and types of QoS parameters for both the physical and network layers are almost same for all the voice, data or video services. But application layer's QoS parameters for different services are far different from each other. Bit Error Rate (BER) and cell coverage are the two most important parameters for physical layer. The performance degradation of these two parameters can degrade the performances of many other parameters. Similarly handoff success ratio for the network layer is the most important parameter in the relationship models. The increase in handoff success ratio reduces the dropped call rate. No user expects end of a call before its call lifetime. These three models are useful for QoS management to increase the QoS parameters performances of heterogeneous 3G networks.

### 3.1 Relationship for Voice Type Applications

Figure 1 shows a relationship model that summarizes the relationship among the QoS parameters which are related to voice type applications. In the application layer, dropped call rate and speech quality are the two most important parameters from the user point of view. The speech quality measured by R value [5] depends on end-to-end delay, delay variation, packet loss rate, codec delay, quality during call duration time, SNR etc. Hence degradation of anyone of these QoS parameters is sufficient to degrade the speech quality.

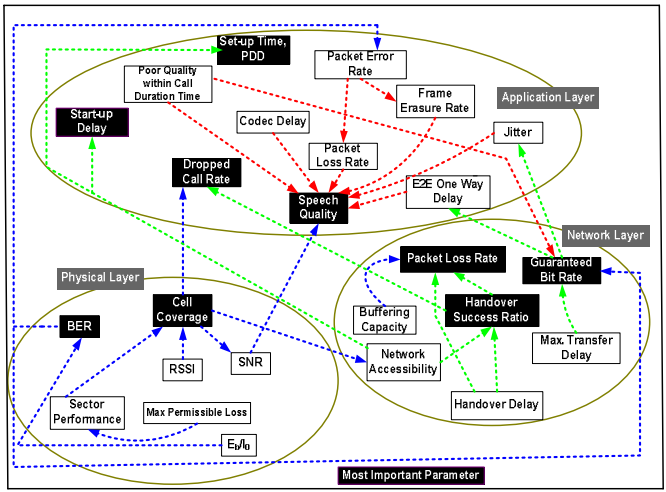

**Figure 1. QoS Relationship Model for Voice Type Applications**

### 3.2 Relationship for Video Type Applications

The relationship model shown in Figure 2 summarizes the relationship among the QoS parameters which are related to video type applications. In the application layer, dropped call rate and resolution are the two most important parameters. The resolution quality depends on frame erasure rate, packet loss rate, codec delay, quality during call duration time, SNR etc. So, for the better quality of resolution, all the related parameters must have better performances. Two other important QoS parameters call set-up delay and start-up delay are influenced by network accessibility.

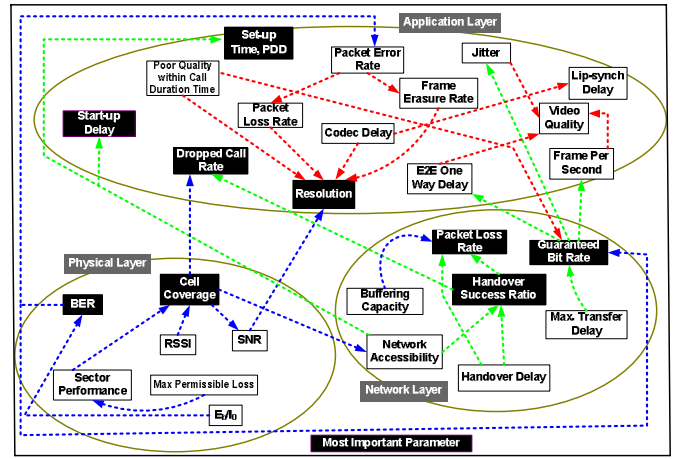

**Figure 2. QoS Relationship Model for Video Type Applications**

### 3.3 Relationship for Data Type Applications

Figure 3 shows the relationship model for data type applications. In the application layer, dropped call rate and data throughput are the two most important parameters. The user satisfaction level mostly depends on the data throughput. Also packet loss is very important for data type applications. This packet loss is directly or indirectly related to BER and packet error rate.

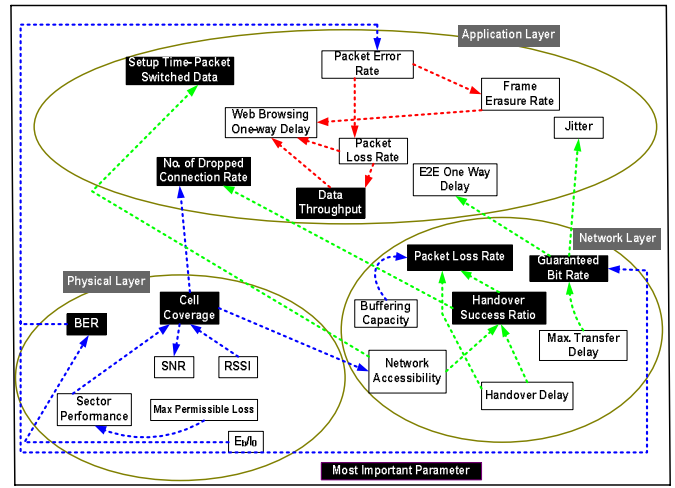

**Figure 3. QoS Relationship Model for Data Type Applications**

## 4. Proposed Soft-QoS Scheme

The proposed soft-QoS scheme to reduce dropped call rate is based on critical bandwidth ratio $\xi_m$ [8]. This critical bandwidth ratio is defined as the ratio of user's minimum bandwidth requirement and the allocated bandwidth. In our proposed scheme we also used two other bandwidth ratios $\xi'_m$ and $\xi_{mp}$. $\xi'_m$ represents the bounded minimum value of bandwidth ratio up to which it can be reduced to accept a new call and $\xi_{mp}$ represents the bandwidth ratio of the existing calls before accepting a call for a traffic class *of m.* The value of $\xi_{mp}$ can be changed from $\xi_m$ to 1. The value of $\xi'_m$ is greater than $\xi_m$ except the real-time conversational traffic class. For the real-time conversational traffic class, both $\xi'_m$ and $\xi_m$ are equal to 1.

Suppose $N_m$ is the total number of calls for $m^{th}$ class while $M$ is the total number of classes and $C_{mN_m}$ is the requested bandwidth for $N_m^{th}$ call. It is possible to calculate the maximum amount of releasable bandwidth $C_{releasable}$ that can be released from the existing calls.

The releasable bandwidth to accept a handoff call is:

$$C_{releasable} = \sum_{m=1}^{M}\sum_{n=1}^{N_m} C_{mn}(\xi_{mp} - \xi_m) \quad (1)$$

The releasable bandwidth to accept a new call is:

$$C_{releasable} = \sum_{m=1}^{M}\sum_{n=1}^{N_m} C_{mn}(\xi_{mp} - \xi'_m) \quad (2)$$

The bandwidth which is already occupied by the existing calls can be calculated as:

$$C_{occupied} = \sum_{m=1}^{M}\sum_{n=1}^{N_m} C_{mn}\xi_{mp} \quad (3)$$

Equations (1) and (2) show that, as $\xi_m$ is less than $\xi'_m$, more bandwidth is releasable to accept a handoff call than a new call.

So, a call will be accepted by a network only when the required bandwidth for that call is less than or equal to this releasable bandwidth. The Call Admission Control (CAC) for the proposed soft-QoS scheme is shown in Figure 4.

This CAC decides whether a call can be accepted or not. After a call request, the CAC checks total capacity $C$, available resources, call type (new or handoff), requested bandwidth $C_{m(N_m+1)}$ for a call and compares the values of bandwidth ratio $\xi_m$, $\xi'_m$ with $\xi_{mp}$.

The limit of minimum bandwidth ratio $\xi'_m$ to release bandwidth form existing calls to accept a new call helps not to reduce the $\xi_{mp}$ value due to accepted new calls. If the value of $\xi_{mp}$ is reduced by accepting new calls, then the existing calls can not release sufficient bandwidth to accept a handoff call. Thus the dropped call rate will not reduce effectively when a user move from one cell to other cell.

## 5. Numerical Result

The numerical results shown in Figures 5 and 6 compare the effect of our soft-QoS scheme with hard QoS mechanism. Table 2 shows the basic parameters which we used for our analysis and the other parameters are as like [8]. The call request sequence is assumed as voice, web traffic, background, voice, video, web traffic, voice, and background. So if total number of requested call is 30, then among them number of voice call, video call, web traffic call and background call will be 11, 4, 8, 7, respectively.

**Table 2. Parameters for Four Traffic Classes**

| Traffic Class (m) | Requested Bandwidth, $C_{mn}$ [Kbps] | $\xi_m$ | $\xi'_m$ |
|---|---|---|---|
| Conversational Voice (m=1) | 16 | 1 | 1 |
| Streaming Video(m=2) | 32 | 0.7 | 0.8 |
| Interactive Web browsing (m=3) | 10 | 0.7 | 0.8 |
| Background (m=4) | 25 | 0.4 | 0.6 |

Figures 5 and 6 show that, after 32 requested calls, hard QoS scheme can accept calls only if some calls complete their call lifetime. Figure 6 shows the reduction of dropped call rate due to proposed soft-QoS scheme. Figure 5 shows that, up to 52 requested new calls, blocked call rate is zero for soft-QoS scheme. But after 62 requested new calls, blocked call rate for soft-QoS scheme is almost same as hard QoS scheme due to heavy traffic condition and at this time handoff calls are more important than new calls.

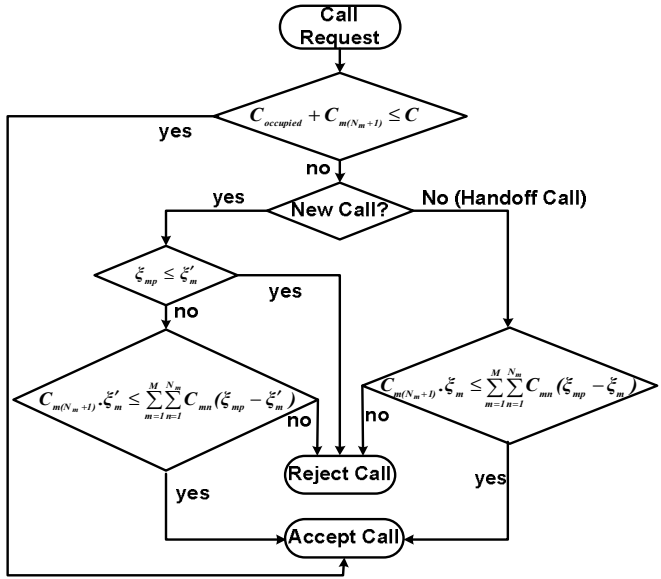

**Figure 4. CAC for Proposed Soft-QoS Scheme**

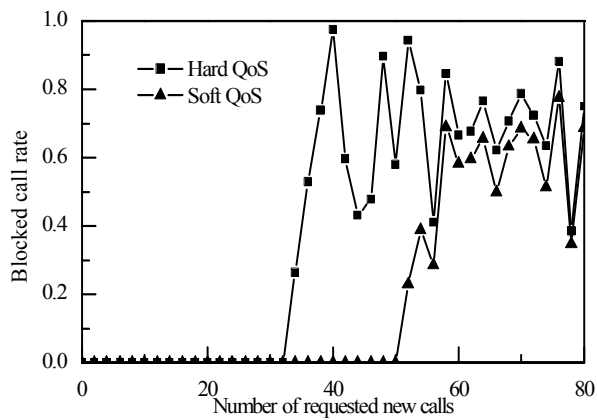

**Figure 5. Blocked Call Rate vs. Number of Requested New Calls**

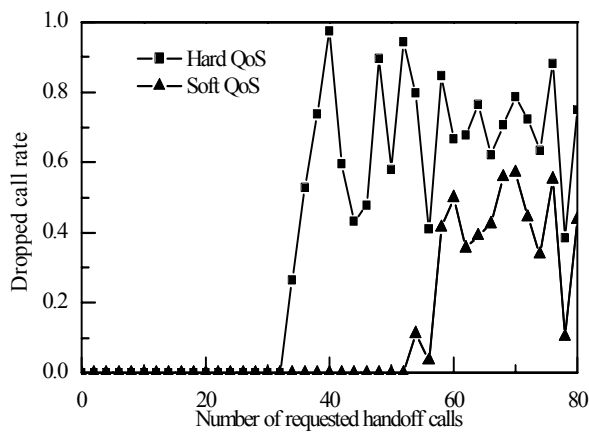

**Figure 6. Dropped Call Rate vs. Number of Requested Handoff Calls**

Hence results indicate that, the proposed soft-QoS scheme can reduce not only dropped call rate but also blocked call rate for medium traffic condition. To accommodate more handoff calls in heavy traffic condition, it gives less priority for new calls than handoff calls.

## 6. Conclusion

This paper summarizes most of the QoS parameters which are related to 3G wireless networks. The QoS parameters with priority order can be helpful for both the operators and users to maximize the network performances and user satisfaction level with the limited resources. During the limited resource condition, QoS requirement can be optimized according to the service type, price, user requirement and priority of QoS parameters. Hence for proper QoS management, this QoS parameter characterization will be very helpful. The QoS relationship models are very helpful to know about the relationship among the QoS parameters which influence other parameters. Thus our proposed models can be useful for efficient management of QoS parameters.

The proposed soft-QoS scheme is found to be very much effective to reduce the dropped call rate which is the most important parameter for all types of services. This scheme can also reduce blocked call rate up to medium traffic condition if bandwidth is not sufficient to accept a requested call.

## Acknowledgement

This research was supported by the MIC (Ministry of Information and Communication), Korea, under the ITRC (Information Technology Research Center) support program supervised by the IITA (Institute of Information Technology Advancement) (IITA-2007-C1090-0603-0019). This work was also supported by the 2007 research fund of Kookmin University and Kookmin University research center UICRC in Korea and KT, Korea.